\documentstyle[aps,prl,multicol,epsf]{revtex}
\title{\bf Kinetics of Particles Adsorption Processes Driven by
Diffusion.}
\author{P. Wojtaszczyk, J.B. Avalos and J.M. Rubi \\Departament de
F\'{\i}sica Fonamental, 
Facultat de F\'{\i}sica,\\Universitat de Barcelona,
\\Avda. Diagonal 647, E-08028 Barcelona (Spain)}
\begin{document}
\maketitle\parskip 2ex
\renewcommand{\theequation}{\arabic{equation}}

\begin{abstract}
The kinetics of the deposition of colloidal particles onto a solid
surface is analytically studied. We take into account both the diffusion
of particles from the bulk as well as the geometrical aspects of the layer
of adsorbed particles. We derive the first kinetic equation for the
coverage of the surface (a generalized Langmuir equation) whose
predictions are in agreement with recent simulation results where diffusion of
particles from
the bulk is explicitly considered. 
\end{abstract}

PACS{68.10.jY,82.70.Dd,05.70.Ln}

\begin{multicols}{2}
\setcounter{equation}{0}        

        The adsorption of colloidal particles onto adsorbing
surfaces is a very common phenomenon in many fields of 
Biology and Chemical-Physics. The deposition of bacteria on teeth is an
example of such a process. There are many other systems in which
suspensions are in contact with adsorbing solid surfaces. The adsorption
of latex spheres onto a flat solid surface is a simple experimental model
system, on which considerable work has been done.  An optical microscope
permits the direct observation of structural properties of the adsorbed
layer\cite{Woj1,Adam}.

        This apparently simple phenomenon involves several fundamental
processes that 
govern the kinetic behavior of the system and the structure of the
adsorbed layer.  First, the attractive potential that characterize the
adsorbing surface plays an essential role.  Depending on the strength of
this potential, adsorbed particles can either desorb or move laterally or
remain irreversibly adsorbed. Second, the kinetics of the adsorption is
strongly influenced by particle transport from the bulk to the
surface.  It depends, for instance, on whether gravity or diffusion
dominates the dynamics of the particles in the bulk. Third, the adsorbing
particles interact with the adsorbed ones in the vicinity of the
surface.  This interaction is repulsive for stable colloidal dispersions and,
therefore, it is responsible for the saturation of the adsorbing surface,
the so-called {\em blocking effect}.

        Theories have been proposed to analyze the contribution of these
effects to the kinetics of the adsorption of colloidal particles. A very
common model is the {\em random sequential adsorption} (RSA)
\cite{rsa}, introduced to generate the structure of a layer
of irreversibly adsorbed colloidal particles. In this model, the adsorbing
particles are deposited 
sequentially at random. The overlaps are forbiden so that, when in a trial
an adsorbing particle overlaps an adsorbed one, this trial is rejected and
one proceeds with a new one.  Clearly, RSA takes the excluded volume
effect into account but it cannot deal with the transport process of the
particles from the bulk to the surface. However, a kinetic equation for
the RSA model has been proposed\cite{Scha2}. If one defines
the coverage of a two-dimensional surface $\theta$ as $\pi R^2 \rho_s$,
$\rho_s$ being the number of adsorbed particles per unit area, and $R$ the
radius of the particles, one then writes
\begin{equation}
\frac{\partial\theta}{\partial t} = K_a \rho_B \Phi_{RSA}(\theta). 
                \label{filang}
\end{equation}
\noindent Here $K_a$ is a kinetic coefficient, $\rho_B$ is the
concentration in the bulk and $\Phi_{RSA}(\theta)$ is the available area at
a given coverage for a 
new particle to adsorbe. 
$\Phi_{RSA}$ was first obtained by Widom\cite{Wid}. 

        Some effort
has also been made in constructing a model incorporating a
description of the transport of adsorbing particles from the bulk. For
spherical particles, heuristic arguments allowed Shaaf {\em et al.}\cite
{Scha} to describe the kinetics of irreversible adsorption for a system
near surface
saturation. When diffusion is explicitly considered, these authors found
that the time-dependence
of the surface coverage is changed with respect to that obtained from the
pure RSA model (eq. (\ref{filang})).  Tarjus and Viot\cite{Tar} later
proposed an analytic (1+1)-dimensional model, in which hard disks were
adsorbed onto a line, in order to give more insight on the previous
results of Schaaf {\em et al}. From ref.\cite{Tar}, one can
infer that the shape of the particles plays a significan role in the
time-dependence of the coverage near saturation.  Currently, for three
dimensional systems the kinetics of adsorption over all time ranges is
only known from simulations\cite{Sen}

        In this letter, we propose a model taking into account both the
diffusion of the particles from the bulk and the saturation of the
surface due to the blocking effect. Our aim is primarily to derive a
kinetic equation ({\em generalized Langmuir equation}) describing the
time-evolution of the coverage for a three-dimensional system. Our
equation is valid not only near saturation but also at intermediate times,
when the surface is only partially covered. The results issued from our
work are in good agreement with recent simulations\cite{Sen} (Fig.1).  Our
secondary aim is to show that eq. (\ref{filang}) corresponds to the
adsorption of diffusing cylinders. This confirms previous authors'claims
as to the crucial role played by the shape of the particles in the
kinetics of adsorption.

        The following analysis can be applied to a large class of
systems, provided that they meet the following requirements; (i) the
attractive potential, responsible for the adsorption of the particles on
the surface, is large enough so that the adsorption can be regarded as
irreversible. Then the particles are not observed to desorb or to
diffuse laterally on the surface; (ii) the surface-particle as well as
the particle-particle potential is of short range, much smaller than
the size of the particles; (iii) diffusion dominates over gravitational
effects in the dynamics of the adsorbing particles, thus the P\'eclet
number (P\'e= mgR/kT, where $m$ is the mass of the particles, $g$ is the
acceleration of the gravity and $kT$ is the thermal energy) of the system
is much smaller than the unity. 

        The procedure adopted here is to consider that the system of
adsorbing particles is in local equilibrium, so that a {\em local}
chemical potential can be defined\cite{Dia,Giv,Ign}. Here, the local
equilibrium hypothesis is assumed since Brownian motion
permits the particles to explore large regions in space before they get
adsorbed. Thus, one expects that the overall adsorption process is not
determined by the local inhomogeneities of the layer of adsorbed particles
but by its global properties, in the spirit of a mean field approach.
Finally, for the system under scrutiny, the only relevant direction will
be that orthogonal to the plane (the $z$-direction) since adsoption will
be responsible for an overall density gradient depending only on $z$.

        Therefore, considering that the suspension of the adsorbing
particles is dilute, one can write 
the local chemical potential as
\begin{equation}
\mu(\gamma) = kT \ln[\rho(\gamma)] - kT\ln\Phi(\theta,\gamma)) 
                \label{mu}
\end{equation}
\noindent where $\gamma \equiv z/2R$ is the dimensionless coordinate in
the $z$-direction, $R$ being the radius of the
particles. The origin of coordinates is thus taken at a distance $R$ 
from the adsorbing plane so that $z=0$ coincides with
the centers of the adsorbed particles. Therefore, it is in the
region $0\leq \gamma \leq 1$ that the interaction between the adsorbing
and the adsorbed particles takes place. This region will be referred to
as the {\em interaction layer} from now on.  In eq.  
(\ref{mu}), the first term in
the right hand side comes from the configurational contribution as in an
ideal gas. The second term accounts for the accessible area at
a given height $\gamma$ and at a given coverage $\theta$, through the
available surface function $\Phi(\theta,\gamma)$, which comes
from an average over the configurations of the deposited particles at a
fixed coverage\cite{Scha3}. Note that, for a
given coverage, the change of the available surface function with $\gamma$
is only due to the shape of the particles. 
It is easy to see that the area excluded at $\gamma$ by a single
adsorbed sphere is simply $4\pi R^2 (1-\gamma^2)$.  Thus, the
available surface function can approximately be written as a function of one
argument, according to $\Phi(\theta,\gamma) \simeq \Phi(\theta
(1-\gamma^2))$. In this approximation an additional
explicit dependence in $\gamma$ is ignored, due to the
fact that in performing the average over the configurations the
true radius of the particles is not the scaled one at a given $\gamma$.
This explicit dependence is not too important for our purposes since the
kinetics mainly depends on the behavior of $\Phi(\theta, \gamma)$ for
$\gamma$ near $0$, which 
corresponds to the bottle-neck for the whole process.
We will further assume that the chemical potential of the bulk
corresponds to that of a dilute suspension of particles,
$\mu_B = kT \ln[\rho_B]$.

Recent experimental results\cite{Woj} indicate that the geometrical
aspects of the adsorbed layer for real systems at small P\'{e}clet number
are practically indistinguishable from those obtained from RSA filling
rules.  Therefore, this allows us to further assume that
$\Phi(\theta,\gamma) \simeq \Phi_{RSA}(\theta (1-\gamma^2))$. Using the
uniform approximation for the $\theta$-dependence of $\Phi_{RSA}$ as given
in ref.\cite{Jin}, we can finally write

\end{multicols}

\begin{equation}
\Phi(\theta,\gamma) = \frac{(1-x(1-\gamma^2))^3}{1 - 0.812 x(1-\gamma^2) +
0.234 x^2(1-\gamma^2)^2 + 0.084 x^3(1-\gamma^2)^3} 
\label{pad}
\end{equation}

\begin{multicols}{2}

\noindent where $x \equiv \theta/\theta_{\infty}$, $\theta_{\infty}$ being
the coverage at saturation.

        To proceed further, we will analyze the flux of particles going to
the surface. Adsorbing particles satisfy a continuity equation of the form
\begin{equation}
\frac{\partial}{\partial t} \rho(\gamma, t)= -\frac{\partial}{\partial
\gamma} J(\gamma,t)     \label{con}
\end{equation}
\noindent where we have considered that the variables depend only on the
$z$-direction through $\gamma$, and on the time. Since the gradient of the
chemical potential (\ref{mu}) is the thermodynamic force responsible
for the flux of particles, one can write 
\begin{equation}
J(\gamma) = - L(\gamma) \frac{\partial \mu(\gamma)}{\partial \gamma}
        \label{1}
\end{equation}
\noindent where we have assumed a linear relation between the flux
$J(\gamma)$ and the thermodynamic force $\partial \mu(\gamma)/\partial
\gamma$, $L(\gamma)$ being a phenomenological coefficient in the spirit
of Onsager's approach\cite{dGr}. In this way, from eqs. (\ref{con}) and
(\ref{1}) we arrive to a Smoluchowski equation for the density of the
adsorbing particles inside the interaction layer
\begin{equation}
\frac{\partial}{\partial t} \rho= \frac{\partial}{\partial
\gamma} \frac{D}{R^2} \left[\frac{\partial}{\partial \gamma}
\rho-\rho \frac{\partial}{\partial \gamma} \ln\Phi \right]       \label{2}
\end{equation}
\noindent where we have defined the {\em diffusion coefficient} $D \equiv
kT LR^2/\rho$ assumed as constant from now on.  The adsorbed particles
induce an entropic barrier, which depends on their shape and on the
coverage, through which the adsorbing ones have to diffuse to reach the
wall. Crucial is to note that eq. (\ref{2}) differs from the Smoluchowski
equation obtained by Shaaf {\em et al} \cite{Scha}. The cross-sectional
area in ref\cite{Scha} is defined only near saturation and the master
equation used have no meaning out of this limit.  Thus the presence of
$\Phi$ in the Smoluchowski equation (\ref{2}) is not trivial.

We will now derive the kinetic equation. We assume that the process is
quasistationary $\partial\rho/\partial t \simeq 0$.  This means that
$J(\gamma)$ is a constant, standing for the fact that no accumlation of
adsorbing particles in the interaction layer occurs.  Thus, mass
conservation leads to $J(\gamma) = J_s$, where $J_s$ is the flux of
particles reaching the adsorbing surface.  We consider here that the
density of particles in the bulk is the control parameter and thus express
$J_s$ in terms of $\rho_B$, with the boundary conditions
\begin{eqnarray}
\mu(\gamma = 1) &=& \mu_B, \label{3}\\
\mu(\gamma = 0) &\rightarrow& -\infty \label{4}
\end{eqnarray}
\noindent The first boundary condition assumes that the chemical potential is 
continuous through the upper boundary of the interaction layer.  The
second stands for an irreversible adsorption process. It is
equivalent to imposing
$\rho(\gamma = 0) = 0$, according to eq.(\ref{mu}), meaning that there are
no free particles in contact with the surface.
In this way, we can
obtain the flux of particles reaching the surface in terms of $\rho_B$ by
solving the differential equation
\begin{equation} 
J_s=-\frac{D}{R^2} \left[\frac{\partial}{\partial \gamma}
\rho-\rho \frac{\partial}{\partial \gamma} \ln \Phi \right] \label{5}
\end{equation}
\noindent with boundary conditions specified in eqs. (\ref{3})
and (\ref{4}). We then arrive at
\begin{equation}
J_s = -\frac{D}{R^2} \rho_B I(\theta) \label{jsu2}
\end{equation}
\noindent where
\begin{equation}
I(\theta)=\frac {1}{\int_{0}^{1}\frac{1}{\Phi(\theta,\gamma)}d\gamma}
\label{jsu}
\end{equation}
\noindent Therefore, the equation for the density of
the adsorbed particles has the form 
\begin{equation}
\frac{\partial\rho_s}{\partial t} =-J_s= \frac{D}{R^2} \rho_B I(\theta)
\label{lang2} 
\end{equation}
\noindent This equation can be rewritten in terms of the coverage by
multiplying both sides by $\pi R^2$, yielding 
a generalized Langmuir equation
\begin{equation}
\frac{\partial\theta}{\partial t} = K_a \rho_B I(\theta) \label{lang}
\end{equation}
\noindent where we have defined the kinetic coefficient $K_a = D \pi$.
Therefore, the main result of this letter is a equation describing the
covering of an adsorbing surface when diffusion as well as the blocking
effect are important. 
 
        The asymptotic time-dependence near saturation is obtained by
analyzing the behavior of $I(\theta)$ when $\theta \rightarrow
\theta_{\infty}$.
In this limit, this function behaves as $I(\theta) \sim
(\theta-\theta_{\infty})^{5/2}$. Inserting this assymptotic behavior in
eq. (\ref{lang}) we find that $\theta -
\theta_{\infty} \sim t^{-2/3}$. Therefore, our equation recovers
the power law behavior near saturation \cite{Scha}.
However, we can go beyond the calculations of ref.\cite{Scha}
because eq.  (\ref{lang}) gives the evolution of the coverage in all
the time range. We have to precise at this point that at the very begining
of the deposition process, the quasistationary conditions break down.
Thus, our analysis is valid
only after this initial time range which is of the order of $R^2/D$,
In particular, for all values of $\theta/\theta_{\infty}$ we find a
remarkable agreement between $I(\theta)$ obtained here and the results
reported in ref.\cite{Sen}, from simulations of
diffusing spheres (Fig. 1). The underlying physics leading the kinetics of
the system is independent of the distance $h$ at which one place the
particle reservoir keeping constant $\rho_B$. However, the quasistationary
condition critically depend on this distance due to the fact that the
density profile relaxes with a time scale $h^2/D$. Note, however, that in
ref.\cite{Sen} the reservoir is located at a distance of the order of $R$,
being then justified the comparison of these simulations with our
calculations.

         Let us now
consider a system in which the particles are cylinders 
(radius $R$ and heigth $2R$) that diffuse with their axis
perpendicular to the adsorbing surface.  In this case, the available
surface function $\Phi$ is independent of $\gamma$. Thus, from eq.
(\ref{jsu}), one straightforwardly gets $I(\theta) =
\Phi_{RSA}(\theta)$. Therefore, for this system we have found the same
kinetic equation as that proposed for the RSA model (eq. (\ref{1})).
Obviously, the assymptotic behavior for the
coverage near saturation for diffusing cylinders gives the RSA power
law decay $\theta-\theta_{\infty} \sim t^{-1/2}$.
Particles of different shape can also be studied. In particular, one can
consider revolution bodies 
wich offer an available surface function of the form
$\Phi(\theta(1-\gamma^{2n})^{1/n})$, which would correspond to spheres for
$n=1$, and to cylinders for $n\rightarrow \infty$. For intermediate values
of $n$, these particles look like cylinders with round edges.
One can then determine the kinetic equations for such objets and find the
corresponding power law behavior near saturation. The assymptotic
behavior depends on $n$ and has the form $\theta-\theta_{\infty}
\sim t^{2n/(1-4n)}$.

       Our analysis, therefore,
gives an interpretation of the previous simulation results
\cite{Sen}, pointing out the collective effect of diffusion, blocking
effect and the shape of the particles, on the kinetics of adsorption. Due
to the generality of the basic statements underlying our theory, one
can envisage its use to the study of the kinetics of more complex systems
as, for instance, suspensions of
elongated particles, provided that the structural
properties of the adsorbed layer are given.

\section*{Acknowledgements}

P. Wojtaszczyk wishes to thank the European Community for the fellowship
of the {\em TMR} program (contract no. ERBFMGECT950062). This work has been
partially supported by DGICyT ({\em Acci\'on Integrada 107B with France}).
The authors are indebted to I.  Pagonabarraga, J. Bafaluy and G. Gomila
for many fruitful discussions and suggestions.

\end{multicols}

\end{document}